\documentclass[showpacs,prd]{revtex4}

\usepackage{epsfig}
\usepackage{graphicx}
\usepackage{dcolumn}
\usepackage{amsmath}
\usepackage{latexsym}

\begin{document}

%%%%%%%%%%%%%%%%%%%%
\title{Non-commutativity from exact renormalization group dualities}  
%\date{\today}
\author{Sunandan Gangopadhyay$^{a,b,c}$\footnote{e-mail:sunandan.gangopadhyay@gmail.com}, 
Frederik G Scholtz $^{a,d}$\footnote{e-mail:
fgs@sun.ac.za}}
\affiliation{$^a$National Institute for Theoretical Physics (NITheP), 
Stellenbosch 7602, South Africa\\
$^b$Department of Physics, West Bengal State University, 
Barasat, Kolkata 700126, India\\
$^c$Inter University Centre for Astronomy and Astrophysics (IUCAA), Pune 411007, India\\
$^d$Institute of Theoretical Physics, 
Stellenbosch University, Stellenbosch 7602, South Africa}

%%%%%%%%%%%%%%%%
\begin{abstract}
\noindent Here we demonstrate, firstly, the construction of dualities using the exact renormalization group approach and, secondly, that spatial non-commutativity  can emerge as such a duality.  This is done in a simple quantum mechanical setting that establishes an exact duality between the commutative and non-commutative quantum Hall systems with harmonic interactions.  It is also demonstrated that this link can be understood as a blocking (coarse graining) transformation in time that relates commutative and non-commutative degrees of freedom.  
%%%%%%%%%%%%%%%%

\end{abstract}

\maketitle

%%%%%%%%%%%%%%%%%%%%%%%%%%%%%%%%%%%

%%%%%%%%%%%%%%%%%%%%%%%%%%%%%%%%%%%%%%%%%%%%%%%%%%%%%%%%%%%%

The premise of the renormalization group as introduced by Wilson \cite{ken} is the description of physical systems at different length scales through effective actions obtained by a transformation of decimation and rescaling \cite{bog}.   The scope of the renormalization group idea was broadened with the introduction of the functional or exact renormalization group that combines functional and renormalization group ideas.  The first versions of the exact renormalization group equation (ERGE) were formulated a long time ago by Wilson \cite{wilson},\cite{kogut} and Wegner\cite{wegner}.  Thereafter, a smooth cutoff version of the ERGE, known as the Polchinski equation \cite{pol}, was derived in order to study field theories. The central idea involves the introduction of an ultra-violet cutoff function $K^{-1}(p^2/\ell^2)$ in the theory.  This real function has the property that it vanishes when $|p|>\ell$. The equation is then obtained by requiring that the process of reducing the number of degrees of freedom (commonly known as the coarse graining step) leaves the normalised generating functional invariant.  An important ingredient in this derivation, required to preserve the form of the source terms,  is the imposition of the condition that the sources vanish above the momentum cutoff and that the cutoff function is independent of the cutoff at small momenta, i.e, $J(p)=0$ for $|p|>\ell$ and $\partial_{\ell}K^{-1}(p^2/\ell^2)=0$ for small $p$ \cite{pol},\cite{banks} . In essence, but completely in the spirit of the renormalization group, this implies that the effective theory can only yield information on correlation functions of the original theory in as far as they are computed below the momentum cutoff.   The ERG (see \cite{bag} for a detailed list of references) has proved to be useful in attacking a large class of problems ranging from relativistic quantum field theories \cite{berges} and statistical systems near the critical point \cite{wett} to the strongly coupled regime of quantum chromodynamics \cite{wett1}. 

It is natural to enquire whether the approach above can be extended by relaxing the conditions imposed on the sources and cutoff function and only require the invariance of the normalised generating functional under the renormalization group flow.  This will allow the computation of all the correlation functions of the original theory in terms of the correlation functions of the effective theory, thereby establishing a complete duality between them.  This, however, necessitates the flow of the sources together with the interacting part of the action. This point of view has been considered seriously in \cite{ros} to extract correlation functions using the ERGE. Here the motivation for this point of view stems from the form of the path integral actions derived in \cite{sgfgs} and \cite{sgfgs1} for a particle in the non-commutative plane without and with a magnetic field, respectively.  If it is noted that these actions differ from the corresponding commutative ones only in a modification of the kinetic energy term, which assumes a rather simple form when the fields are Fourier transformed, yet do not fulfil the requirements as imposed in the standard ERG, it becomes natural to enquire whether these non-commutative actions may arise naturally from a generalised ERG approach that only relies on invariance of the generating functional. This is the main issue explored here using the simplest, yet non-trivial, quantum mechanical setting of the Landau problem where all computations can be done exactly and analytically.

It should be mentioned that the notion of non-commutative theories as low energy effective theories is in itself not new, but has been explored in quantum Hall systems already some time ago  \cite{dunne}-\cite{fgs}, while the notion that non-commutativity may arise from coarse graining was explored in the context of position-momentum non-commutativity in \cite{wetter}.

 We start by summarising the essentials of the ERGE.  For our present purposes it is sufficient to consider a complex scalar field theory in 0+1-dimensions, but the generalisation to higher dimensions and theories with different field content is straightforward.  We take the following action :
\begin{eqnarray}
\label{action}
S[\phi, \phi^{*}]=\int d\omega~\phi^{*}(\omega)K(\omega, \ell)\phi(\omega)+S_{I}[\phi, \phi^{*}]+J_{\ell}[\phi, \phi^{*}]
\label{1}
\end{eqnarray}
where $K(\omega, \ell)$ takes the standard form $\omega^2$ in the $\ell\rightarrow0$ limit and $J_{\ell}[\phi, \phi^{*}]$ is a generalised source term, which is a functional of the fields, determined by the requirement of invariance of the generating functional.   For our present purposes, which only involves actions quadratic in the fields,  it is simple to see that it is sufficient to limit the form of this functional to be linear, i.e., we take
\begin{eqnarray}
J_{\ell}[\phi, \phi^{*}]=\int d\omega~[J_{0}(\ell) +J^{*}_{0}(\ell) +J_{1}(\ell)\phi^{*}(\omega)+J^{*}_{1}(\ell)\phi(\omega)],
\label{2}
\end{eqnarray}
and impose the initial conditions
\begin{eqnarray}
\label{2x}
J_{0}(\ell)|_{\ell=0}&=&J^{*}_{0}(\ell)|_{\ell=0}=0 \\
J_{1}(\ell)|_{\ell=0}=J_{1}(0)&,&J^{*}_{1}(\ell)|_{\ell=0}=J^{*}_{1}(0).
\label{2y}
\end{eqnarray}
Here $J(0)$ is an arbitrary function of $\omega$ that acts as a source in the bare ($\ell=0$) action.   In what follows we denote the first term $\phi^{*}(\omega)K(\omega, \ell)\phi(\omega)$ in eq.(\ref{action}) by $S_{0}[\phi, \phi^{*}]$, and we refer to the second term $S_{I}[\phi, \phi^{*}]$ as the interacting part of the action.  

The normalised generating functional is given by
\begin{eqnarray}
Z[J_{\ell}]=\frac{\int [d\phi~d\phi^{*}]~e^{-(S_{0}[\phi, \phi^{*}] +S_{I}[\phi, \phi^{*}]+J_{\ell}[\phi, \phi^{*}])}}{\int [d\phi~d\phi^{*}]~e^{-(S_{0}[\phi, \phi^{*}] +S_{I}[\phi, \phi^{*}])}}~.
\label{3}
\end{eqnarray}
We now apply the logic of ERGE as set out in \cite{pol} and \cite{banks}, by requiring this to be invariant under the flow, i.e., it must be independent of $\ell$.  However, we relax the conditions imposed in these derivations on $K(\omega, \ell)$ and the sources, which necessitates the flow of the source terms.  A derivation similar to the one carried out in \cite{banks} yields the following equations for the interacting part and source terms:
\begin{eqnarray}
\label{9}
\partial_{\ell}S_{I}=\int d\omega~\partial_{\ell}K^{-1}
\left\{\frac{\delta S_{I}}{\delta \phi^{*}(\omega)}\frac{\delta S_{I}}{\delta \phi(\omega)}
-\frac{\delta^{2} S_{I}}{\delta \phi^{*}(\omega)\delta \phi(\omega)}\right\}
\end{eqnarray}
\begin{eqnarray}
\partial_{\ell}J_{\ell}&=&\int d\omega~\partial_{\ell}K^{-1}
\left\{\frac{\delta S_{I}}{\delta \phi(\omega)}\frac{\delta J_{\ell}}{\delta \phi^{*}(\omega)}
+\frac{\delta S_{I}}{\delta \phi^{*}(\omega)}\frac{\delta J_{\ell}}{\delta \phi(\omega)}
+\frac{\delta J_{\ell}}{\delta \phi^{*}(\omega)}\frac{\delta J_{\ell}}{\delta \phi(\omega)}
-\frac{\delta^{2} J_{\ell}}{\delta \phi^{*}(\omega)\delta \phi(\omega)}\right\}.
\label{10}
\end{eqnarray}
These equations can easily be solved when the interaction term is quadratic in the fields, i.e., 
\begin{eqnarray}
S_{I}[\phi, \phi^{*}]=\int d\omega~g(\omega, \ell) \phi^{*}(\omega) \phi(\omega)
\label{11}
\end{eqnarray}
with $g(\omega, \ell)$ real.  Focussing on the source terms for the moment and using eqs.(\ref{2}) and (\ref{10}) yields
%\begin{eqnarray}
%\int d\omega~\left[(\partial_{\ell}J_{0}(\ell)+c.c.)+(\partial_{\ell}J_{1}(\ell)\phi^{*}(\omega)+c.c.)\right]\nonumber\\
%=\int d\omega~\partial_{\ell}K^{-1}
%[g(\omega, \ell)(\phi^{*}(\omega)J_{1}(\ell)
 %+c.c.)+J^{*}_{1}(\ell)J_{1}(\ell)]\nonumber\\
%\label{12}
%\end{eqnarray}
%where c.c. denotes complex conjugate. The above equation implies
\begin{eqnarray}
\label{13a}
\partial_{\ell}J_{1}(\ell)&=&\partial_{\ell}K^{-1}(\omega, \ell)g(\omega, \ell)  J_{1}(\ell),\\
\label{13b}
\partial_{\ell}J^{*}_{1}(\ell)&=&\partial_{\ell}K^{-1}(\omega, \ell)g(\omega, \ell) J^{*}_{1}(\ell),\\
\partial_{\ell}[J_{0}(\ell)+J^{*}_{0}(\ell)]&=&\partial_{\ell}K^{-1}(\omega, \ell)|J_{1}(\ell)|^{2}.
\label{13c}
\end{eqnarray}
Integrating these equations and using the initial conditions (\ref{2x}) and (\ref{2y}) on the sources, we obtain
\begin{eqnarray}
\label{14a}
J_{1}(\ell)&=&J_{1}(0)\exp\left(\int_{0}^{\ell} d\ell'~g(\omega, \ell')\partial_{\ell'}K^{-1}(\omega, \ell')\right)\\
\label{14b}
J^{*}_{1}(\ell)&=&J^{*}_{1}(0)\exp\left(\int_{0}^{\ell} d\ell'~g(\omega, \ell')\partial_{\ell'}K^{-1}(\omega, \ell')\right)\\
J_{0}(\ell)+J^{*}_{0}(\ell)&=&|J_{1}(0)|^{2}\int_{0}^{\ell} d\ell'~\partial_{\ell'}K^{-1}(\omega, \ell')
\exp\left(2\int_{0}^{\ell'} d\ell''~g(\omega, \ell'')\partial_{\ell''}K^{-1}(\omega, \ell'')\right).
\label{14c}
\end{eqnarray}
We now apply these results to the Landau problem.  The action for a particle moving in a plane with a constant magnetic field perpendicular to the plane reads, in complex coordinates $z=x+iy$,
\begin{eqnarray}
S=\int d\omega~[m\bar{z}(\omega)\omega^2 z(\omega) +eB\omega\bar{z}(\omega)z(\omega) 
+J(\omega)\bar{z}(\omega)+J^{*}(\omega)z(\omega)].
\label{15}
\end{eqnarray}
We now flow this action with respect to a parameter $\ell$ as in eq.(\ref{1}), using eq.(\ref{11}) for the interaction and taking
%\begin{eqnarray}
%S=\int d\omega~[\bar{z}(\omega)K(\omega, \ell) z(\omega) +g(\omega, \ell)\bar{z}(\omega)z(\omega)]+J_{\ell}[z, \bar z]\nonumber\\
%\label{16}
%\end{eqnarray}
%where we choose $K(\omega, \ell)$ to be
\begin{eqnarray}
K(\omega, \ell)=\frac{m\omega^2}{(1-\frac{m\omega\ell}{\hbar})},
\label{17}
\end{eqnarray}
with the initial condition
\begin{eqnarray}
g(\omega, \ell)|_{\ell=0}=eB\omega.
\label{18}
\end{eqnarray}
The choice of $K(\omega, \ell)$ is 
motivated by the non-local form of the kinetic energy term appearing in the action for a particle moving in a non-commutative plane \cite{sgfgs}.  
Substituting the form (\ref{11}) of $S_{I}$ in eq.(\ref{9}), we obtain the following flow equation for the coefficient $g(\omega, \ell)$ :
\begin{eqnarray}
\frac{\partial g(\omega, \ell)}{\partial\ell}=-\frac{1}{\hbar\omega}g^{2}(\omega, \ell).
\label{19}
\end{eqnarray}
Note that we have ignored a vacuum term in the above equation which originates from the second term on the right hand side
of eq.(\ref{9}).
Integration of this equation subject to the initial condition (\ref{18}) yields
\begin{eqnarray}
g(\omega, \ell)=e\tilde{B}(\ell)\omega~;~~\tilde{B}(\ell)=\frac{B}{(1+\frac{eB\ell}{\hbar})}.
\label{20}
\end{eqnarray}
Using these results, we obtain the effective action as
\begin{eqnarray}
S=\int d\omega\left[\bar{z}(\omega)K(\omega, \ell)z(\omega)+e\tilde{B}(\ell)\omega\bar{z}(\omega)z(\omega)\right]+J_{\ell}[z, \bar{z}].\nonumber\\
\label{21}
\end{eqnarray}
We now recast this form of the action in the following way (which brings it closer to the form of the action for
a particle in a magnetic field moving on a noncommutative plane) :
\begin{eqnarray}
S&=&\int d\omega\left[\bar{z}(\omega)K(\omega, \ell)\left(1-\frac{e\tilde{B}(\ell)}{\hbar}\right)z(\omega)
+\frac{e\tilde{B}(\ell)\omega}{(1-\frac{m\omega\ell}{\hbar})}\bar{z}(\omega)z(\omega)\right]+J_{\ell}[z, \bar{z}].
\label{21x}
\end{eqnarray}
Rescaling the coordinates as
\begin{eqnarray}
\tilde{z}(\omega)=\sqrt{1-\frac{e\tilde{B}(\ell)\ell}{\hbar}}~z(\omega),
\label{23}
\end{eqnarray}
using the identity $\tilde{B}(\ell)/(1-\frac{e\tilde{B}(\ell)\ell}{\hbar})=B$ and the following form of the sources obtained by substituting eqs.(\ref{17}) and (\ref{20}) in eqs.(\ref{14a}, \ref{14b}, \ref{14c}):
\begin{eqnarray}
\label{26a}
J_{1}(\ell)&=&\frac{J_{1}(0)}{(1+\frac{eB\ell}{\hbar})}\\
\label{26b}
J^{*}_{1}(\ell)&=&\frac{J^{*}_{1}(0)}{(1+\frac{eB\ell}{\hbar})}\\
J_{0}(\ell)+J^{*}_{0}(\ell)&=&-\frac{|J_{1}(0)|^{2}\ell}{\hbar\omega(1+\frac{eB\ell}{\hbar})}~,
\label{26c}
\end{eqnarray}
we obtain the final form of the action and sources:
\begin{eqnarray}
S&=&\int d\omega\left[\bar{\tilde z}(\omega)K(\omega, \ell) \tilde{z}(\omega)
 +\frac{eB\omega}{(1-\frac{m\omega\ell}{\hbar})}\bar{\tilde z}(\omega)\tilde z(\omega)
-\frac{|J_{1}(0)|^{2}\ell}{\hbar\omega(1+\frac{eB\ell}{\hbar})}
+\left(\frac{J_{1}(0)}{\sqrt{1+\frac{eB\ell}{\hbar}}}\bar{\tilde{z}}(\omega)+c.c.\right)\right].\nonumber\\
\label{28}
\end{eqnarray}

We have thus constructed a one parameter family of theories all of which have the same physics
as the one represented by eq.(\ref{15}). It is reassuring to note that the generating functionals 
computed from the actions (\ref{15}) and (\ref{28}) agree with each other thereby verifying the internal consistency 
of our analysis.

Comparing with the action of a particle moving in a magnetic field $B^{*}$ in a non-commutative plane with non-commutative parameter $\ell$ \cite{sgfgs1} and source terms:
\begin{eqnarray}
S_{NC}=\int d\omega\left[\bar{\tilde z}(\omega)K(\omega, \ell) \tilde{z}(\omega)
 +\frac{eB^{*}(1+\frac{eB^{*}\ell}{4\hbar})\omega}{(1-\frac{m\omega\ell}{\hbar})}
\bar{\tilde z}(\omega)\tilde z(\omega)+J\bar{\tilde{z}}(\omega) +\bar{J}\tilde{z}(\omega)\right]
\label{29}
\end{eqnarray}
establishes a complete duality between these two systems if the following identification between the magnetic fields $B$ and $B^{*}$ is made:
\begin{eqnarray}
B=B^{*}\left(1+\frac{eB^{*}\ell}{4\hbar}\right).
\label{30}
\end{eqnarray}
The above relation between the magnetic fields in the commutative and non-commutative planes was also obtained earlier in a path integral framework in \cite{sgfgs1}.  The modified source terms appearing in (\ref{28}) simply establishes the precise dictionary between the correlation functions of the two systems. 

We now proceed to demonstrate that this result admits a different interpretation in terms of a blocking procedure \cite{ros}, where the blocking is performed
over time. To see this, recall that it is possible to interpret the renormalization flow as a change of variables in the path integral \cite{ros}.  Bar the source terms this is in our case a rather simple change of variables involving a $\omega$ dependent scaling,
$\tilde{z}(\omega)=\sqrt{1-m\omega\ell/\hbar}~z(\omega)$, that transforms the commutative action (\ref{15}) into the non-commutative action (\ref{28}). It is easy to check that this corresponds to the following blocking relation \cite{ros}
\begin{eqnarray}
\tilde{z}(t)=\int_{-\infty}^{+\infty}~dt'~\frac{e^{-i\omega_{0}(t'-t)}}{\sqrt{\omega_0}(t' -t)^{3/2}}z(t'),
\label{rost}
\end{eqnarray}
where $\omega_0=\frac{\hbar}{m\ell}$.  Some care is required in the prescription of evaluating this integral that manifests itself in divergent dimensionless normalisation factors that need to be regularised, but that are completely independent of any physical parameters.  The important point is that this relation clearly reveals the fact that spatial non-commutativity can be seen as a blocking procedure over time.
It is also reassuring to check, using the commutator of $z(t)$ and $\bar{z}(t')$ ($t\neq t'$)
\begin{eqnarray}
[z(t),\bar{z}(t')]=\frac{2\hbar}{eB}(e^{-ieB (t-t')/m}-1),
\label{uneq}
\end{eqnarray}
that the equal time commutator of $\tilde{z}(t)$ and $\bar{\tilde z}(t)$ is given by $[\tilde{z}(t),\bar{\tilde z}(t)]=c\ell$.  Note, though, that the equal time commutator of $z(t)$ and $\bar{z}(t^\prime)$ vanishes.  The computation of the dimensionless constant $c$ again requires the regularisation of divergent integrals, but is completely independent of any physical parameters, i.e., the commutator is determined by $\ell$ only.  This is consistent with our rationale of mapping the original theory onto a non-commutative theory with non-commutative parameter $\ell$.  In this regard it is important to note that this only happens if the coarse graining relation (\ref{rost}) is used.  

Next we consider the action for a particle in a magnetic field in the presence of a harmonic oscillator potential :
\begin{eqnarray}
S=\int d\omega~[m\bar{z}(\omega)\omega^2 z(\omega) +(eB\omega-k)\bar{z}(\omega)z(\omega) 
+J(\omega)\bar{z}(\omega)+J^{*}(\omega)z(\omega)]
\label{15ha}
\end{eqnarray}
where $k$ is the strength of the harmonic oscillator interaction. We can once again flow this action with respect
to the parameter $\ell$ to the form (\ref{1}) with interaction as in eq.(\ref{11}). To obtain the form of the flow for the coefficients $g(\omega, \ell)$
governed by the flow eq.(\ref{19}), we make the following ansatz :
\begin{eqnarray}
g(\omega, \ell)=\frac{G(\omega, \ell) \omega}{(1-\frac{m\omega\ell}{\hbar})}
\label{ha1}
\end{eqnarray}
where $G(\omega, \ell)$ is an unknown function, which is to be determined, and impose the initial condition
\begin{eqnarray}
g(\omega, \ell)|_{\ell=0}=eB\omega-k ;\implies G(\omega, \ell)|_{\ell=0}=eB-\frac{k}{\omega}.
\label{ha2}
\end{eqnarray}
Substituting this ansatz in eq.(\ref{19}), we obtain the following equation for $G(\omega, \ell)$ :
\begin{eqnarray}
\frac{\partial G(\omega, \ell)}{\partial \ell}+\frac{m\omega}{\hbar(1-\frac{m\omega\ell}{\hbar})}
G(\omega, \ell)=- \frac{1}{\hbar(1-\frac{m\omega\ell}{\hbar})}G^2(\omega, \ell).\nonumber\\
\label{ha4}
\end{eqnarray}
Integrating this equation subject to the initial condition (\ref{ha2}) leads to
\begin{eqnarray}
G(\omega, \ell)=\hbar F(\omega, \ell)\left(1-\frac{m\omega\ell}{\hbar}\right)~;~
F(\omega, \ell)=\frac{eB\omega-k}{[\hbar\omega +(eB\omega-k)\ell]}\nonumber\\
\label{ha5}
\end{eqnarray}
which in turn implies (using eq.(\ref{17}))
\begin{eqnarray}
g(\omega, \ell)=\frac{\hbar\omega F(\omega, \ell)}{(1-\frac{m\omega\ell}{\hbar})}
-K(\omega, \ell)F(\omega, \ell)\ell.
\label{ha7}
\end{eqnarray}
Using these results and rearranging the terms, we obtain the effective action as
\begin{eqnarray}
S=\int d\omega\left[\bar{\tilde z}(\omega)K(\omega, \ell)\tilde{z}(\omega)
+\frac{eB\omega-k}{(1-\frac{m\omega\ell}{\hbar})}\bar{\tilde z}(\omega)\tilde z(\omega)\right]
+J_{\ell}[\tilde z, \bar{\tilde{z}}]
\label{ha8}
\end{eqnarray}
where we have  rescaled the coordinates as follows:
\begin{eqnarray}
\tilde{z}(\omega)=\sqrt{1-F(\omega, \ell)\ell}~z(\omega).
\label{ha9}
\end{eqnarray}
Note that this rescaling is $\omega$ dependent in contrast to the earlier case (\ref{23}) in the absence of the harmonic oscillator potential.

Using $-\frac{k}{(1-\frac{m\omega\ell}{\hbar})}
=-k -\frac{mk\omega\ell}{\hbar(1-\frac{m\omega\ell}{\hbar})}$, we can recast the above action as
\begin{eqnarray}
S=\int d\omega\left[\bar{\tilde z}(\omega)K(\omega, \ell)\tilde{z}(\omega)
+\frac{(eB-mk\ell/\hbar)\omega}{(1-\frac{m\omega\ell}{\hbar})}\bar{\tilde z}(\omega)\tilde z(\omega)
-k\bar{\tilde z}(\omega)\tilde{z}(\omega)\right]
+J_{\ell}[\tilde z, \bar{\tilde{z}}].
\label{ha10}
\end{eqnarray}
The above form clearly indicates that the magnetic field has been renormalized by the harmonic oscillator interaction strength.
Further, the form of the action is also identical to the action of a particle in a magnetic field in the presence of a harmonic
oscillator potential moving on a non-commutative plane \cite{sgfgs1}.

The flows of the sources can also be obtained easily by substituting
eqs.(\ref{17}) and (\ref{ha7}) in eqs.(\ref{14a}, \ref{14b}, \ref{14c}) to yield
\begin{eqnarray}
\label{ha11}
J_{1}(\ell)&=&J_{1}(0)\frac{\hbar\omega}{[\hbar\omega+(eB\omega-k)\ell]}\\
\label{ha12}
J^{*}_{1}(\ell)&=&J^{*}_{1}(0)\frac{\hbar\omega}{[\hbar\omega+(eB\omega-k)\ell]}\\
J_{0}(\ell)+J^{*}_{0}(\ell)&=&-\frac{|J_{1}(0)|^{2}\ell}{[\hbar\omega+(eB\omega-k)\ell]}~.
\label{ha13}
\end{eqnarray}
The above equations together with eq.(\ref{ha10}) constitutes a one parameter family of theories dual to the one of eq.(\ref{15ha}).  Again, the flow of the source terms establishes the precise dictionary between the correlation functions of the two systems.

We now summarize our findings. In this paper, we have constructed one parameter families of dual theories using the ERGE for the Landau problem without and with a harmonic oscillator potential.  The precise choice made here for the kinetic energy term, as captured by the function $K(\omega,\ell)$, established these dual families to be non-commutative theories. 
The form for the flow of the sources has also been computed in both cases.
We also observe a subtle link between non-commutative theories and the blocking procedure in the ERG approach.  The latter opens up an avenue to explore these relations in higher dimensional and interacting field theories and will be pursued elsewhere.  
\vskip 1 cm

\noindent {\bf{Acknowledgements}}: This work was supported under a grant of the National Research Foundation of South Africa. 

%%%%%%%%%%%%%%%%%%%%%%%%%%%%%%%%%%%%%%%%%%%%%%%%%%%%%%%%%%%%%%%%%%%%%%%%%%%

%%%%%%%%%%%%%%%%%%%%%%%%%%%%%%%%%%%%%%%%%%%%%%%%%%%%%%%%%%%%%%%%%%%%%%%%%%%%
%%%%%%%%%%%%%%%%%%%%%%%%%%%%%%%%%%%%%%%%%%%%%%%%%%%%%%%%%%%%%%%%%%%%%%%%%%%%
\end{document}